\documentclass[twocolumn,showpacs,preprintnumbers,amsmath,amssymb,pra]{revtex4}

\usepackage{graphicx}
\usepackage{dcolumn}
\usepackage{bm}

\begin{document}

\title{An optical clock with neutral atoms confined in a shallow trap}

 \author{Pierre Lemonde$^1$}
 \email{pierre.lemonde@obspm.fr}
\author{Peter Wolf$^{1,2}$} \affiliation{$^1$ SYRTE, Observatoire de
Paris\\ 61, Avenue de l'observatoire, 75014, Paris, France\\ $^2$
Bureau International des Poids et Mesures,\\ Pavillon de
Breteuil,\\ 92312 S{\`e}vres Cedex, France}

\date{\today}

\begin{abstract}
We study the trap depth requirement for the realization of an
optical clock using atoms confined in a lattice. We show that
site-to-site tunnelling leads to a residual sensitivity to the
atom dynamics hence requiring large depths ($50$ to $100\,E_r$ for
Sr) to avoid any frequency shift or line broadening of the atomic
transition at the $10^{-17}-10^{-18}$ level. Such large depths and
the corresponding laser power may, however, lead to difficulties
(e.g. higher order light shifts, two-photon ionization, technical
difficulties) and therefore one would like to operate the clock in
much shallower traps. To circumvent this problem we propose the
use of an accelerated lattice. Acceleration lifts the degeneracy
between adjacents potential wells which strongly inhibits
tunnelling. We show that using the Earth's gravity, much shallower
traps (down to $5\,E_r$ for Sr) can be used for the same accuracy
goal.
\end{abstract}

\maketitle

\section{Introduction\label{sec:intro}}

The control of the external degrees of freedom of atoms, ions and
molecules and of the associated frequency shifts and line
broadenings is a long standing issue of the fields of spectroscopy
and atomic frequency standards. They have been a strong motivation
for the development of many widely spread techniques like the use
of buffer gases\,\cite{Dicke53}, Ramsey
spectroscopy\,\cite{Ramsey85}, saturated
spectroscopy\,\cite{lee67}, two-photon
spectroscopy\,\cite{Biraben74}, trapping and laser
cooling\,\cite{Leibfried03,CCTNobel}, etc.

In the case of ions, the problem is now essentially solved since
they can be trapped in relatively low fields and cooled to the
zero point of motion of such traps\,\cite{Leibfried03}. In this
state, the ions are well within the Lamb-Dicke
regime\,\cite{Dicke53} and experience no recoil nor first order
Doppler effect\,\cite{Leibfried03}. The fractional accuracy of
today's best ion clocks lies in the range from 3 to 10$\times
10^{-15}$\,\cite{Berkeland98,Udem01,Stenger01_2,Madej04,Margolis04}
with still room for improvement. The main drawback of trapped ion
frequency standards is that only one to a few ions can contribute
to the signal due to Coulomb repulsion. This fundamentally limits
the frequency stability of these systems and puts stringent
constraints on the frequency noise of the oscillator which probes
the ions\,\cite{Young99}.

These constraints are relaxed when using a large number of neutral
atoms\,\cite{Quessada03} for which, however, trapping requires
much higher fields, leading to shifts of the atomic levels. This
fact has for a long time prevented the use of trapped atoms for
the realization of atomic clocks and today's most accurate
standards use freely falling atoms. Microwave fountains now have
an accuracy slightly below $10^{-15}$ and are coming close to
their foreseen ultimate limit which lies around
$10^{-16}$\,\cite{Bize04}, which is essentially not related to
effects due to the atomic dynamics\,\cite{Wolf04,Li04}. In the
optical domain, atomic motion is a problem and even with the use
of ultra-cold atoms probed in a Ramsey-Bord\'{e}
interferometer\,\cite{Borde84}, optical clocks with neutrals still
suffer from the first order Doppler and recoil
effects\,\cite{Ishikawa94,Borde02,Sterr04,Oates05}. Their
state-of-the-art accuracy is about $10^{-14}$\,\cite{Sterr04}.

The situation has recently changed with the proposal of the
optical lattice clock\,\cite{KatoPal03}. The idea is to engineer a
lattice of optical traps in such a way that the dipole potential
is exactly identical for both states of the clock transition,
independently of the dipole laser power and polarisation. This is
achieved by tuning the trap laser to the so-called "magic
wavelength" and by the choice of clock levels with zero electronic
angular momentum. The original scheme was proposed for $^{87}$Sr
atoms using the strongly forbidden $^1S_0-^3P_0$ line at 698\,nm
as a clock transition\,\cite{Courtillot03}. In principle however,
it also works for all atoms with a similar level structure like
Mg, Ca, Yb, Hg, etc. including their bosonic isotopes if one uses
multi-photon excitation of the clock
transition\,\cite{Hong05,Santra04}.

In this paper we study the effect of the atom dynamics in the
lattice on the clock performances. In ref.\,\cite{KatoPal03}, it
is implicitly assumed that each microtrap can be treated
separately as a quadratic potential in which case the situation is
very similar to the trapped ion case and then fully
understood\,\cite{Leibfried03}. With an accuracy goal in the
$10^{-17}-10^{-18}$ range in mind (corresponding to the mHz level
in the optical domain), we shall see later on, that this is
correct at very high trap depths only. The natural energy unit for
the trap dynamics is the recoil energy associated with the
absorption or emission of a photon of the lattice laser,
$E_r=\frac{\hbar^2k_L^2}{2m_a}$ with $k_L$ the wave vector of the
lattice laser and $m_a$ the atomic mass. For Sr and for the above
accuracy goal the trap depth $U_0$ corresponding to the
independent trap limit is typically $U_0=100\,E_r$, which
corresponds to a peak laser intensity of 25\,kW/cm$^2$.

For a number of reasons however, one would like to work with traps
as shallow as possible. First, the residual shift by the trapping
light of the clock transition is smaller and smaller at a
decreasing trap depth. The first order perturbation is
intrinsically cancelled by tuning to the magic wavelength except
for a small eventual tensorial effect which depends on the
hyperfine structure of the atom under consideration. Higher order
terms may be much more problematic depending on possible
coincidences between two photon resonances and the magic
wavelength\,\cite{KatoPal03,Porsev04}. The associated shift scales
as $U_0^2$\,\footnote{note that this effect cannot be quantified
without an accurate knowledge of the magic wavelength and of the
strength of transitions involving highly excited states.}. The
shifts would then be minimized by a reduction of $U_0$ and its
evaluation would be greatly improved if one can vary this
parameter over a broader range. Second, for some of the possible
candidate atoms, such as Hg for which the magic wavelength is
about 340 nm, two-photon ionization can occur which may limit the
achievable resonance width and lead to a frequency shift. Finally,
technical aspects like the required laser power at the magic
wavelength can be greatly relaxed if one can use shallow traps.
This can make the experiment feasible or not if the magic
wavelength is in a region of the spectrum where no readily
available high power laser exists, such as in the case of Hg. For
this atom, a trap depth of $100\,E_r$ would necessitate a peak
intensity of $500\,$kW/cm$^{2}$ at 340\,nm.

When considering shallow traps, the independent trap limit no
longer holds, and one cannot neglect tunnelling of the atoms from
one site of the lattice to another. This leads to a delocalization
of the atoms and to a band structure in their energy spectrum and
associated dynamics. In section \ref{sec:periodic} we investigate
the ultimate performance of the clock taking this effect into
account. We show that depending on the initial state of the atoms
in the lattice, one faces a broadening and/or a shift of the
atomic transition of the order of the width of the lowest energy
band of the system. For Sr, this requires $U_0$ of the order of
$100\,E_r$ to ensure a fractional accuracy better than $10^{-17}$.

The deep reason for such a large required value of $U_0$ is that
site-to-site tunnelling is a resonant process in a lattice. We
show in section \ref{sec:accelerated} that a much lower $U_0$ can
be used provided the tunnelling process is made non-resonant by
lifting the degeneracy between adjacent sites. This can be done by
adding a constant acceleration to the lattice, leading to the
well-known Wannier-Stark ladder of
states\,\cite{Nenciu91,Gluck98}. More specifically, we study the
case where this acceleration is simply the Earth's gravity. The
experimental realization of the scheme in this case is then
extremely simple: the atoms have to be probed with a laser beam
which propagates vertically. In this configuration, trap depths
down to $U_0\sim 5\,E_r$ can be sufficient for the above accuracy
goal.

\section{Confined atoms coupled to a light field\label{sec:confined}}

In this section we describe the theoretical frame used to investigate the
residual effects of the motion of atoms in an external potential.
The internal atomic structure is approximated by a two-level
system $|g\rangle$ and $|e\rangle$ with energy difference
$\hbar\omega_{eg}$. The internal Hamiltonian is:
\begin{equation}
    \hat{H}_{i}=\hbar\omega_{eg}|e\rangle \langle e|.
\end{equation}
We introduce the coupling between $|e\rangle$ and $|g\rangle$ by a
laser of frequency $\omega$ and wavevector $k_s$ propagating along
the $x$ direction:
\begin{equation}
    \hat{H}_{s}=\hbar\Omega \cos(\omega t-k_s\hat{x})|e\rangle \langle
    g|+h.c.,
\end{equation}
with $\Omega$ the Rabi frequency.

In the following we consider external potentials induced by trap
lasers tuned at the magic wavelength and/or by gravity. The
external potential $\hat{H}_{ext}$ is then identical for both
$|g\rangle$ and $|e\rangle$ with eigenstates $|m\rangle$ obeying
$\hat{H}_{ext}|m\rangle=\hbar \omega_m | m\rangle$ (Note that
$|m\rangle$ can be a continuous variable in which case the
discrete sums in the following are replaced by integrals). If we
restrict ourselves to experiments much shorter than the lifetime
of state $|e\rangle$ (for $^{87}$Sr, the lifetime of the lowest
$^3P_0$ state is 100 s) spontaneous emission can be neglected and
the evolution of the general atomic state
\begin{equation}
    |\psi_{at}\rangle = \sum_m a_m^g \,e^{-i\omega_m t}\,|m,g\rangle + a_m^e
\,e^{-i(\omega_{eg}+\omega_m)t}\,|m,e\rangle
\end{equation}
is driven by
\begin{equation}
    i\hbar \frac{\partial}{\partial t}
    |\psi_{at}\rangle=(\hat{H}_{ext}+\hat{H}_{i}+\hat{H}_{s})|\psi_{at}\rangle ,
\end{equation}
leading to the following set of coupled equations
\begin{eqnarray}\label{eq:aevolution}
  i\, \dot{a}_m^g&=&\sum_{m'} \frac{\Omega^*}{2} e^{i\Delta_{m',m}t}\langle m|e^{-ik_s\hat{x}}| m'\rangle a_{m'}^e \\
  \nonumber
  i\, \dot{a}_m^e&=&\sum_{m'} \frac{\Omega}{2} e^{-i\Delta_{m,m'}t}\langle m|e^{ik_s\hat{x}}| m'\rangle
  a_{m'}^g.
\end{eqnarray}
To derive eq. (\ref{eq:aevolution}) we have made the usual rotating
wave approximation (assuming $\omega-\omega_{eg}<<\omega_{eg}$)
and defined $\Delta_{m',m}=\omega-\omega_{eg}+\omega_m-\omega_{m'}$.

In the case of free atoms, $\hat{H}_{ext}=\frac{\hbar^2
\hat{\kappa}^2}{2 m_{a}}$ with $\hbar \hat{\vec{\kappa}}$ the
atomic momentum and $m_a$ the atomic mass. The eigenstates are
then plane waves: $|g,\vec{\kappa}\rangle$ is coupled to
$|e,\vec{\kappa}+\vec{k}_s\rangle$ with
$\Delta_{\vec{\kappa},\vec{\kappa}+\vec{k}_s}=
\omega-\omega_{eg}+\frac{\hbar
\vec{\kappa}.\vec{k}_s}{m_a}+\frac{\hbar k_s^2}{2m_a}$. One
recovers the first order Doppler and recoil frequency shifts.

Conversely in a tightly confining trap $\langle m
|e^{ik_s\hat{x}}| m'\neq m\rangle<< \langle m|e^{ik_s\hat{x}}|
m\rangle$, and the spectrum of the system consists of a set of
unshifted resonances corresponding to each state of the external
hamiltonian. Motional effects then reduce to the line pulling of
these resonances by small (detuned) sidebands\,\cite{Leibfried03}.

\section{Periodic potential\label{sec:periodic}}

\subsection{Eigenstates and coupling by the probe laser\label{sec:eigenstates}}

We now consider the case of atoms trapped in an optical lattice.
As is clear from eq. (\ref{eq:aevolution}), only the motion of the
atoms along the probe laser propagation axis plays a role in the
problem and we restrict the analysis to 1D\footnote{See section
\ref{sec:conclusion} for a brief discussion of the 3D problem}. We
assume that the lattice is formed by a standing wave leading to
the following external hamiltonian:
\begin{equation}
    \hat{H}_{ext}^I=\frac{\hbar^2 \hat{\kappa}^2}{2
m_{a}}+\frac{U_0}{2}(1-\cos(2k_l\hat{x})),
\end{equation}
where $k_l$ is the wave vector of the trap laser. The eigenstates
$|n,q\rangle$ and eigenenergies $\hbar\omega_{n,q}$ of
$\hat{H}_{ext}^I$ are derived from the Bloch
theorem\,\cite{Ashcroft76}. They are labelled by two quantum
numbers: the band index $n$ and the quasi-momentum $q$.
Furthermore they are periodic functions of $q$ with period $2k_l$
and the usual convention is to restrict oneself to the first
Brillouin zone $q\in ]-k_l,k_l]$.

\begin{figure}
    \includegraphics[angle=270,width=8.5 cm]{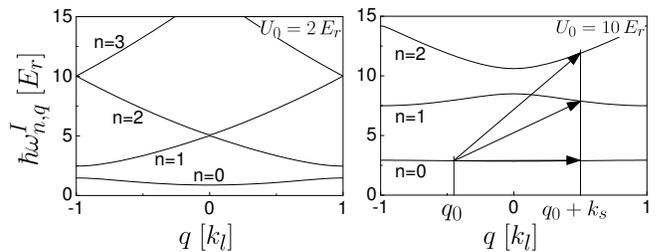}
\caption{Band structure for two different lattice depth:
$U_0=2\,E_r$ (left) and $U_0=10\,E_r$ (right). Each state
$|n,q_0\rangle$ is coupled to all the states $|n',q_0+k_s\rangle$
by the probe laser.}\label{fig:bands}
\end{figure}

\begin{figure}
  \includegraphics[angle=270,width=8.5 cm]{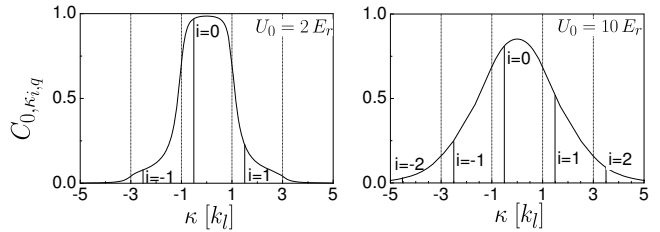}
\caption{$C_{0,\kappa_{i,q}}$ for two different lattice depth:
$U_0=2\,E_r$ (left) and $U_0=10\,E_r$ (right). The bold vertical
lines illustrate the case $q=-k_l/2$. The dotted lines delimit the
Brillouin zones. For a state $|n=0,q=ak_l\rangle$ with $a\in
]-1,1]$ the solid envelope gives the contribution of the plane
waves $|\kappa_{i,ak_l}=ak_l+2ik_l\rangle$.}\label{fig:c0}
\end{figure}

\begin{figure}
  \includegraphics[angle=270,width=8.5 cm]{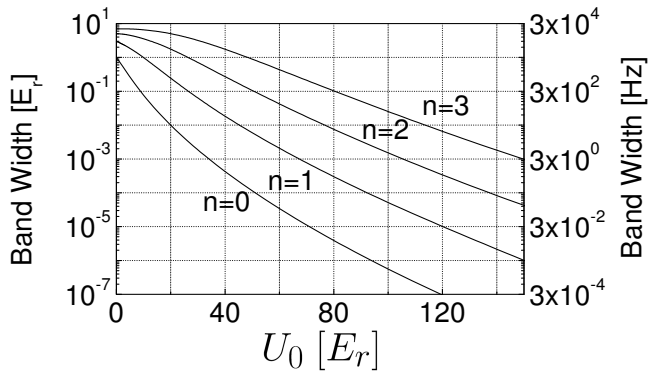}
\caption{Lowest four band widths as a function of the lattice
depth $U_0$ in units of $E_r$ (left scale) and in frequency units
(right scale).}\label{fig:bandwidth}
\end{figure}

Following a procedure given in Ref.\,\cite{Gluck98} a numerical
solution to this eigenvalue problem can be easily found in the
momentum representation. The atomic plane wave with wave vector
$\kappa$ obeys
\begin{equation}\label{eq:H0}
    \hat{H}_{ext}^I|\kappa\rangle=\left(\frac{\hbar^2 \kappa^2}{2
m_{a}}+\frac{U_0}{2}\right)|\kappa\rangle-\frac{U_0}{4}(|\kappa+2k_l\rangle+|\kappa-2k_l\rangle).
\end{equation}
For each value of $q$, the problem then reduces to the
diagonalization of a real tridiagonal matrix giving the
eigenenergies and eigenvectors as a linear superposition of plane
waves:
\begin{eqnarray}\nonumber
 \hat{H}_{ext}^I|n,q\rangle&=& \hbar \omega_{n,q}^I|n,q\rangle \\ \label{eq:blochvectors}
  |n,q\rangle &=&
  \sum_{i=-\infty}^{\infty}C_{n,\kappa_{i,q}}|\kappa_{i,q}\rangle,
\end{eqnarray}
with $\kappa_{i,q}=q+2ik_l$. For each value of $q$ one obtains a
discrete set of energies $\hbar \omega_{n,q}^I$ and coefficients
$C_{n,\kappa_{i,q}}$, which are real and normalized such that
$\sum_i C_{n,\kappa_{i,q}}^2=1$. In figures \ref{fig:bands} and
\ref{fig:c0} are shown $\hbar\omega_{n,q}^I$ and
$C_{0,\kappa_{i,q}}$ for various values of $U_0$. Except when
explicitly stated, all numerical values throughout the paper are
given for $^{87}$Sr at a lattice laser wavelength $813$\,nm which
corresponds to the magic wavelength reported in
Ref.\,\cite{Takamoto03}. In frequency units $E_r$ then corresponds
to $3.58$\,kHz. In figure\,\ref{fig:bandwidth} is shown the width
($|\omega_{n,q=k_l}^I-\omega_{n,q=0}^I|$) of the lowest energy
bands as a function of $U_0$ in units of $E_r$ and in frequency
units.


Substituting $\langle m| \rightarrow \langle n,q|$ and $|m'\rangle
\rightarrow |n',q'\rangle$ in eq. (\ref{eq:aevolution}), the
action of the probe laser is described by the coupled equations

\begin{eqnarray}\label{eq:aevolutionlattice}
  i\, \dot{a}_{n,q}^g&=& \sum_{n'}\frac{\Omega^{n',{n}^*}_{q}}{2} e^{i\Delta^{n',n}_{q}t} a_{n',q+k_s}^e \\
  \nonumber
  i\, \dot{a}_{n,q+k_s}^e&=& \sum_{n'}\frac{{\Omega^{n,n'}_{q}}}{2}
  e^{-i\Delta^{n,n'}_{q}t}a_{n',q}^g \hspace{0.3cm},
\end{eqnarray}
with $\Omega^{n,n'}_{q}=\Omega \sum_i
C_{n',\kappa_{i,q}}C_{n,\kappa_{i,q+k_s}}$ and
$\Delta^{n,n'}_{q}=\omega-\omega_{eg}+\omega_{n',q}^I-\omega_{n,q+k_s}^I$.
As expected from the structure of the Bloch vectors in
(\ref{eq:blochvectors}), the translation in momentum space
$e^{ik_s\hat x}$ due to the probe laser leads to the coupling of a
given state $|n,q\rangle$ to the whole set $|n',q+k_s\rangle$ (see
figure \ref{fig:bands}) with a coupling strength
$\Omega^{n',n}_{q}$ and a shift with respect to the atomic
resonance $\omega_{n',q+k_s}^I-\omega_{n,q}^I$. Both quantities
depend on $n$, $n'$ and $q$ and to go further we have to make
assumptions on the initial state of the atoms in the lattice.

\subsection{Discussion\label{sec:discussion1}}

\begin{figure}
\includegraphics[angle=270,width=8.5 cm]{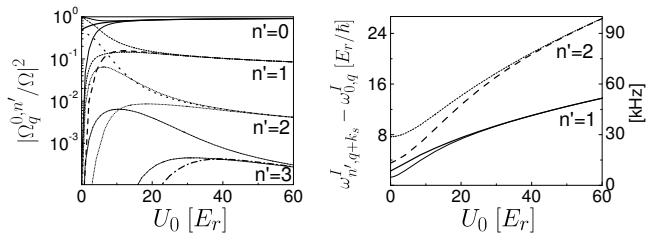}
\caption{Left: Relative strength of the transitions to different
bands ($n=0\rightarrow n'$) for an atom prepared in state
$|n=0,q=-k_l\rangle$ (bold lines), $|n=0,q=-k_l/2\rangle$ and
$|n=0,q=k_l/2\rangle$ (thin lines). Right: detuning of the first
two sidebands for an atom prepared in state $|n=0,q=-k_l\rangle$
(bold lines) and $|n=0,q=0\rangle$ (thin lines) in units of $E_r$
 (left scale) and in frequency units (right scale).}\label{fig:omega_shift_no_gravity}
\end{figure}

\begin{figure}
\includegraphics[angle=270,width=5 cm]{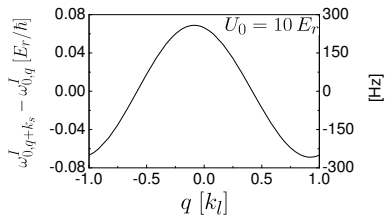}
\caption{Shift of the "carrier" resonance in the first band for a
lattice depth $U_0=10\,E_r$. Left scale: in units of $E_r$. Right
scale: in frequency units.}\label{fig:shift_no_gravity}
\end{figure}

We first consider the case where the initial state is a pure
$|n,q\rangle$ state. The strengths of the resonances
$\Omega^{n,n'}_{q}$ are shown in figure
\ref{fig:omega_shift_no_gravity} for the case $n=0$ and various
values of $q$. At a growing lattice depth $\Omega^{n,n'}_{q}$
become independent of $q$ and the strength of all "sidebands"
($n'-n\neq 0$) asymptotically decreases as $U_0^{-|n'-n|/4}$ for
the benefit of the "carrier" ($n'=n$). The frequency separation of
the resonances rapidly increases with $U_0$ (Fig.
\ref{fig:omega_shift_no_gravity}). For $U_0$ as low as $5\,E_r$,
this separation is of the order of 10\,kHz. For narrow resonances
(which are required for an accurate clock) they can be treated
separately and the effect of the sidebands on the carrier is
negligible. If for example one reaches a carrier width of 10\,Hz,
the sideband pulling is of the order of $10^{-5}$\,Hz. On the
other hand, the "carrier" frequency is shifted from the atomic
frequency by $\omega_{n,q+k_s}^I-\omega_{n,q}^I$ due to the band
structure. This shift is of the order of the width of the
$n^{\text{th}}$ band (Fig. \ref{fig:shift_no_gravity} and
\ref{fig:bandwidth}). It can be seen as a residual Doppler and
recoil effect for atoms trapped in a lattice and is a consequence
of the complete delocalisation of the eigenstates of the system
over the lattice. The "carrier" shift is plotted in figure
\ref{fig:shift_no_gravity} for the case $n=0$ and $U_0=10\,E_r$.
For this shift to be as small as 5\,mHz over the whole lowest
band, which corresponds in fractional units to $10^{-17}$ for Sr
atoms probed on the $^1S_0-^3P_0$ transition, the lattice depth
should be at least $90\,E_r$ (Fig. \ref{fig:bandwidth}).

\begin{figure}
\includegraphics[angle=270,width=5 cm]{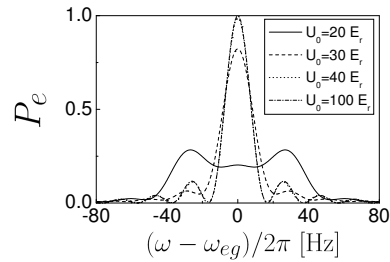}
\caption{Expected resonances in the case where the first band is
uniformely populated for $\Omega =10$\,Hz and
$U_0=20\,E_r,\,30\,E_r,\,40\,E_r,$ and 100\,$E_r$. The duration of
the interaction is such that the transition probability at
resonance is maximized.}\label{fig:resonances_nogravity}
\end{figure}

Another extreme situation is the case where one band is uniformly
populated. In this case the "carrier" shift averaged over $q$
cancels and one can hope to operate the clock at a much lower
$U_0$ than in the previous case. The problem is then the ultimate
linewidth that can be achieved in the system, which is of the
order of the width of the band and is reminiscent of Doppler
broadening. This is illustrated in figure
\ref{fig:resonances_nogravity} for which we have computed the
expected "carrier" resonances in the case where the lowest band is
uniformly populated, by numerically solving equations
(\ref{eq:aevolution}). This was done for a Rabi frequency
$\Omega=10\,$Hz and an interaction duration which is adjusted for
each trapping depth so as to maximize the transition probability
at zero detuning. We have checked that all resonances plotted in
figure \ref{fig:resonances_nogravity} are not shifted to within
the numerical accuracy (less than $10^{-5}$\,Hz). However, at
decreasing $U_0$ the contrast of the resonance starts to drop for
$U_0<40\, E_r$ and the resonance broadens progressively, becoming
unusable for precise spectroscopy when the width of the energy
band reaches the Rabi frequency. To get more physical insight into
this phenomenon, let's consider the particular example of this
uniform band population where one well of the lattice is initially
populated. This corresponds to a given relative phase of the Bloch
states such that the interference of the Bloch vectors is
destructive everywhere except in one well of the lattice. The time
scale for the evolution of this relative phase is the inverse of
the width of the populated energy band which then corresponds to
the tunnelling time towards delocalization (once the relative
phases have evolved significantly, the destructive/constructive
interferences of the initial state no longer hold). The broadening
and loss of contrast shown in figure
\ref{fig:resonances_nogravity} can be seen as the Doppler effect
associated with this tunnelling motion.

The two cases discussed above (pure $|n,q\rangle$ state and
uniform superposition of all states inside a band: $\int dq
|n,q\rangle$) correspond to the two extremes one can obtain when
populating only the bottom band. They illustrate the dilemma one
has to face: either the resonance is affected by a frequency shift
of the order of the width of the bottom band (pure state), or by a
braoadening of the same order (superposition state), or by a
combination of both (general case). In either case the solution is
to increase the trap depth in order to decrease the energy width
of the bottom band.

In the experimental setup described in \cite{Takamoto03} about
90\,\% of the atoms are in the lowest band and can be selected by
an adequate sequence of laser pulses. The residual population of
excited bands can then be made negligible ($<10^{-3}$). On the
other hand, knowing and controlling with accuracy the population
of the various $|q\rangle$ states in the ground band is a
difficult task. The actual initial distribution of atomic states
will lie somewhere between a pure state in the bottom band and a
uniform superposition of all states in the bottom band. If we
assume that the population of the $|q\rangle$ states in the ground
band can be controlled so that the frequency shift averages to
within one tenth of the band width, then a fractional accuracy
goal of $10^{-17}$ implies $U_0=70\, E_r$ or more. Note that due
to the exponential dependence of the width of the ground band on
$U_0$ (see figure \ref{fig:bandwidth}) the required lattice depth
is largely insensitive to an improvement in the control of the
initial state. If for example the averaging effect is improved
down to 1\,\% the depth requirement drops from $70\,E_r$ to
$50\,E_r$. Consequently, operation of an optical lattice clock
requires relatively deep wells and correspondingly high laser
power, which, in turn, is likely to lead to other difficulties as
described in the introduction.

Fortunately, the requirement of deep wells can be significantly
relaxed by adding a constant acceleration to the lattice, as
described in the next section.

\section{Periodic potential in an accelerated frame\label{sec:accelerated}}

\subsection{Wannier-Stark states and coupling by the probe laser\label{sec:eigenstates2}}

\begin{figure}
  \includegraphics[width=5 cm]{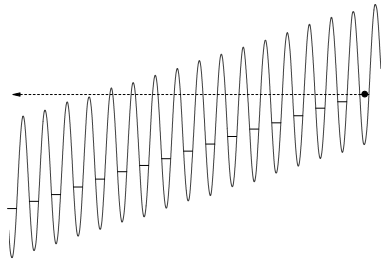}
  \caption{External potential seen by the atoms in the case of
  a vertical lattice ($U_0=5\,E_r$). An atom initially
  trapped in one well of the lattice will end-up in the continuum by tunnel
  effect. For $U_0=5\,E_r$ the lifetime of the quasi-bound state of
  each well is about $10^{10}$\,s.}\label{fig:tiltedlattice}
\end{figure}

\begin{figure}
  \includegraphics[angle=270,width=8.5 cm]{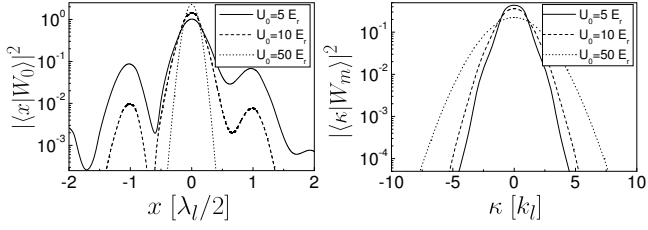}
  \caption{Wannier-Stark states in position (left) and momentum
(right) representation for $U_0=5\,E_r$, $U_0=10\,E_r$ and
$U_0=50\,E_r$. Numerically we first compute the momentum
representation $\langle\kappa|W_m\rangle=b_m(\kappa)C_{0,\kappa}$
and then obtain the position representation by Fourier
tranformation.}\label{fig:wannierstates}
\end{figure}

The shift and broadening encountered in the previous section are
both due to site-to-site tunnelling and to the corresponding
complete delocalization of the eigenstates of the lattice. As is
well-known from solid-state physics, one way to localize the atoms
is to add a linear component to the Hamiltonian\,\cite{Nenciu91}:
adjacent wells are then shifted in energy, which strongly inhibits
tunnelling. In this section we study the case where the lattice
and probe laser are oriented vertically so that gravity plays the
role of this linear component. The external hamiltonian is then:
\begin{equation}
    \hat{H}_{ext}^{II}=\frac{\hbar^2 \hat{\kappa}^2}{2
m_{a}}+\frac{U_0}{2}(1-\cos(2k_l\hat{x}))+m_ag\hat{x},
\end{equation}
with $g$ the acceleration of the Earth's gravity. This hamiltonian
supports no true bound states, as an atom initially confined in
one well of the lattice will end up in the continuum due to
tunnelling under the influence of gravity (Fig.
\ref{fig:tiltedlattice}). This effect is known as Landau-Zener
tunnelling and can be seen as non-adiabatic transitions between
bands induced by the linear potential in the Bloch
representation\,\cite{Zener32,Landau32,Peik97,Bharucha97}. The
timescale for this effect however increases exponentially with the
depth of the lattice and for the cases considered here is orders
of magnitude longer than the duration of the
experiment\,\footnote{This exponential increase is true on average
only and can be modified for specific values of $U_0$ by a
resonant coupling between states in distant
wells\,\cite{Avron82,Bharucha97,Gluck99}.}. In the case of Sr in
an optical lattice, and for $U_0$ as low as $5\,E_r$, the lifetime
of the ground state of each well is about $10^{10}$\,s! The
coupling due to gravity between the ground and excited bands can
therefore be neglected here. In the frame of this approximation
the problem of finding the "eigenstates" of $\hat{H}_{ext}^{II}$
reduces to its diagonalization in a sub-space restricted to the
ground band\,\cite{Wannier60,Callaway63} (we drop the band index
in the following to keep notations as simple as possible). We are
looking for solutions to the eigenvalue equation, of the form:
\begin{eqnarray}\label{eq:eigenproblem_wannierstark}
  \hat{H}_{ext}^{II}|W_m\rangle &=& \hbar \omega_m^{II}|W_m\rangle \\
 \nonumber 
  |W_m\rangle &=& \int_{-k_l}^{k_l}dq\, b_m(q)|q\rangle \,.
\end{eqnarray}
In eq. (\ref{eq:eigenproblem_wannierstark}) the $|q\rangle$ are
the Bloch eigenstates of $\hat H^I_{ext}$ (c.f. section
\ref{sec:periodic}) for the bottom energy band ($n=0$), $m$ is a
new quantum number, and the $b_m(q)$ are periodic:
$b_m(q+2ik_l)=b_m(q)$. After some algebra, eq.
(\ref{eq:eigenproblem_wannierstark}) reduce to the differential
equation
\begin{equation}\label{eq:diff_wannierstark}
    \hbar (\omega_q^I-\omega_m^{II})b_m(q)+im_ag \partial_q b_m(q)=0
\end{equation}
where $\omega_q^I$ is the eigenvalue of the Bloch state
$|n=0,q\rangle$ of section \ref{sec:periodic}. Note that equations
(\ref{eq:eigenproblem_wannierstark}) and
(\ref{eq:diff_wannierstark}) only hold in the limit where
Landau-Zener tunnelling between energy bands is negligible.
Otherwise, terms characterising the contribution of the other
bands must be added and the description of the quasi-bound states
is more complex\,\cite{Avron82,Gluck98}. In our case the
periodicity of $b_m(q)$ and a normalization condition lead to a
simple solution of the form
\begin{eqnarray}
  \omega_m^{II} &=& \omega_0^{II}+m\Delta_g \label{bnq} \\
  b_m(q) &=&
  \frac{1}{\sqrt{2k_l}}e^{-\frac{i\hbar}{m_ag}(q\omega_m^{II}-\gamma_q)} \nonumber
\end{eqnarray}
with the definitions
$\omega_0^{II}=\frac{1}{2k_l}\int_{-k_l}^{k_l}dq\,\omega_q^I$,
$\hbar\Delta_g=m_ag\lambda_l/2$, and
$\partial_q\gamma_q=\omega^I_q$ with $\gamma_0=0$. The
$|W_m\rangle$ states are usually called Wannier-Stark states and
their wave functions are plotted in figure \ref{fig:wannierstates}
for various trap depths. In the position representation
$|W_m\rangle$ exhibits a main peak in the $m^{\textrm{th}}$ well
of the lattice and small revivals in adjacent wells. These
revivals decrease exponentially at increasing lattice depth. At
$U_0=10\,E_r$ the first revival is already a hundred times smaller
than the main peak. Conversely, in the momentum representation,
the distribution gets broader with increasing $U_0$. The phase
shift between $b_m$ and $b_{m-1}$ in (\ref{bnq}), $b_m(q)=e^{-i\pi
q/k_l}b_{m-1}(q)$, corresponds to a translational symmetry of the
Wannier-Stark states in the position representation $\langle
x+\lambda_l/2|W_m\rangle=\langle x|W_{m-1}\rangle$. The discrete
quantum number $m$ is the "well index" characterising the well
containing the main peak of the wave function $\langle
x|W_{m}\rangle$, and, as intuitively expected, the energy
separation between adjacent states is simply the change in
gravitational potential between adjacent wells: $\hbar
\Delta_g=m_ag\lambda_l/2$.

Substituting $\langle m| \rightarrow \langle W_m|$ and $|m'\rangle
\rightarrow |W_{m'}\rangle$ in eq. (\ref{eq:aevolution}) shows
that the effect of the probe laser is to couple the Wannier-Stark
states to their neighbours by the translation in momentum space
$e^{ik_s\hat x}$, with the coupling strengths
\begin{equation}\label{eq:wanniercoupling}
   \langle
   W_m|e^{ik_s\hat{x}}|W_{m'}\rangle=\int_{-\infty}^{\infty}d\kappa \,
   b_m^*(\kappa+k_s)b_{m'}(\kappa)C_{0,\kappa}C_{0,\kappa+k_s},
\end{equation}
obtained from direct substitution of
(\ref{eq:eigenproblem_wannierstark}) and
(\ref{eq:blochvectors})\,\footnote{For similar reasons as in
section\,\ref{sec:periodic} one can neglect the coupling to
excited bands in the system for narrow enough resonances.}.

Using the translational symmetry of the Wannier-Stark states it is easy to show that
\begin{equation}\label{eq:WSsymm}
\langle W_m|e^{ik_s\hat{x}}|W_{m'}\rangle = e^{i\pi m k_s/k_l}\langle
   W_0|e^{ik_s\hat{x}}|W_{m'-m}\rangle.
\end{equation}
From that property, equation (\ref{eq:wanniercoupling}), and using
$b_m(\kappa)=b_m^*(-\kappa)$ (note that $\gamma_q=\gamma_{-q}$)
one can then show that
\begin{equation}\label{eq:symmetrysidebands}
\langle W_m|e^{ik_s\hat{x}}|W_{m+j}\rangle = e^{i\pi j
k_s/k_l}\langle
   W_m|e^{ik_s\hat{x}}|W_{m-j}\rangle,
\end{equation}
which is a useful result when studying the symmetry of coupling to neighbouring states (see next section).

The differential equations (\ref{eq:aevolution}), governing the
evolution of the different states under coupling to the probe
laser are then
\begin{eqnarray}\label{eq:aevolutionwannier}
  i\, \dot{a}_{m}^g&=& \sum_{m'}\frac{\Omega_{m-m'}^*}{2}e^{-i\pi m'\frac{k_s}{k_l}} e^{i\Delta_{m-m'}t} a_{m'}^e \\
  \nonumber
  i\, \dot{a}_{m}^e&=& \sum_{m'}\frac{\Omega_{m'-m}}{2} e^{i\pi m\frac{k_s}{k_l}}
  e^{-i\Delta_{m'-m}t}a_{m'}^g,
\end{eqnarray}
in which we have used (\ref{eq:WSsymm}) and defined $\Omega_m=\Omega \langle
W_0|e^{ik_s\hat{x}}|W_{m}\rangle$ and
$\Delta_m=\omega-\omega_{eg}+m\Delta_g$.

\subsection{Discussion\label{sec:results2}}

\begin{figure}
  \includegraphics[width=6 cm]{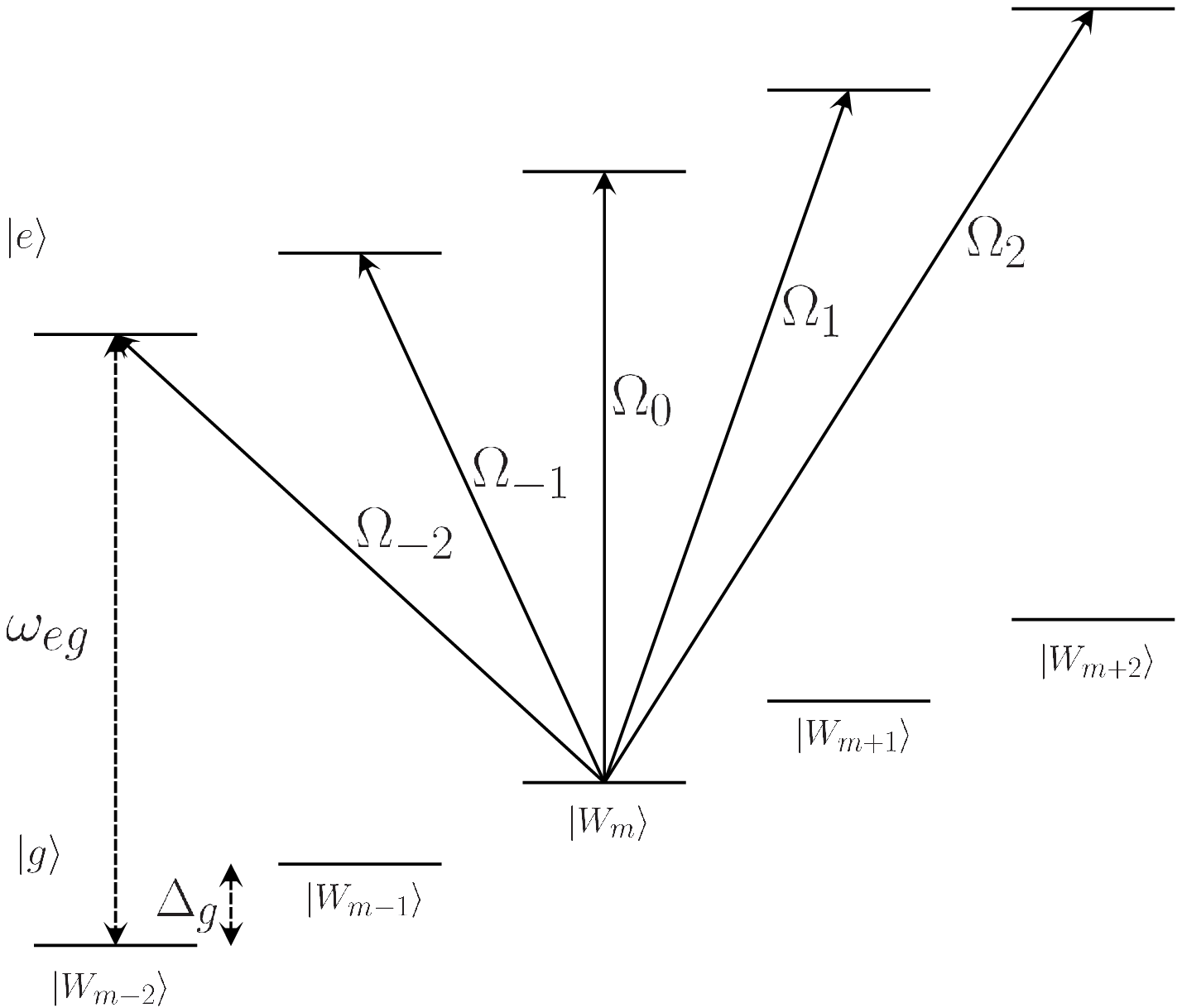}
  \caption{Wannier-Stark ladder of states and coupling between states by the probe laser.
  }\label{fig:wannier_stark_couplings}
\end{figure}

\begin{figure}
  \includegraphics[angle=270,width=8.5 cm]{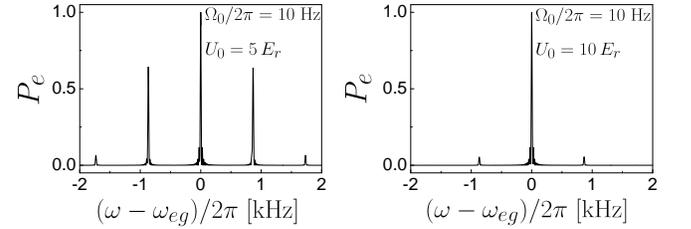}
\caption{Computed resonances when the initial state is a pure
Wannier-Stark state. Left: $U_0=5\,E_r$, right: $U_0=10\,E_r$.
Both resonances are plotted for an effective Rabi frequency of the
carrier $\frac{\Omega_0}{2\pi}=10\,$Hz and for an interaction time
of 50\,ms.}\label{fig:resonances_g}
\end{figure}

\begin{figure}
  \includegraphics[angle=270,width=5 cm]{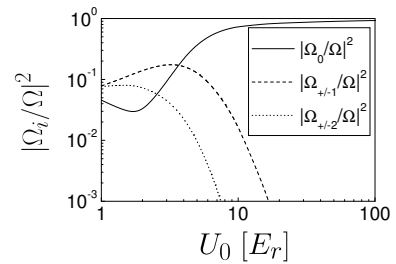}
  \caption{Relative strength of the carrier $|\Omega_0/\Omega|^2$ and of the first
  four sidebands $|\Omega_{\pm 1}/\Omega|^2$ and $|\Omega_{\pm 2}/\Omega|^2$ as a function
  of the lattice depth $U_0$.}\label{fig:omegasgravity}
\end{figure}

We now study the case where the initial state of the atom is a
pure Wannier-Stark state. According to eq.
(\ref{eq:aevolutionwannier}) excitation by the laser will lead to
a set of resonances separated by $\Delta_g$ (see Fig.
\ref{fig:wannier_stark_couplings}). In the case of Sr,
$\Delta_g/2\pi=866\,$Hz and for the narrow resonances required for
high performance clock operation, they are easily resolved. The
resonances obtained by first numerically integrating
(\ref{eq:wanniercoupling}) and then numerically solving
(\ref{eq:aevolutionwannier}) are plotted in figure
\ref{fig:resonances_g} for the cases $U_0=5\,E_r$ and
$U_0=10\,E_r$. They exhibit remarkable properties. First the
"carrier" (which corresponds to the transition
$|W_m\rangle\rightarrow|W_m\rangle$) has a frequency which exactly
matches the atomic frequency $\omega_{eg}$. It also doesn't suffer
from any broadening or contrast limitation (provided the side
bands are resolved) which would be due to the atomic dynamics.
Second, the sidebands ($|W_m\rangle\rightarrow|W_{m\pm i}\rangle$)
have a coupling strength which very rapidly decreases as $U_0$
increases (see fig. \ref{fig:omegasgravity}). In addition they are
fully symmetric with respect to the carrier which results from eq.
(\ref{eq:symmetrysidebands}), and hence lead to no line pulling of
the carrier. We have checked that the numerical calculations agree
with this statement to within the accuracy of the calculations.
This absence of shift and broadening remains true even for very
shallow traps down to a depth of a  few $E_r$, the ultimate
limitation being the lifetime of the Wannier-Stark states. This
situation is in striking contrast with the results of section
\ref{sec:periodic} in the absence of gravity.

\begin{figure}
  \includegraphics[width=5 cm]{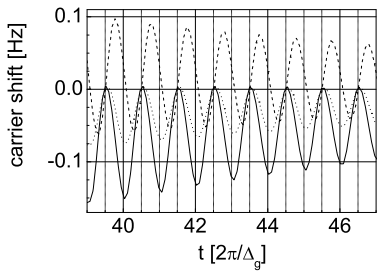}
\caption{Frequency shift of the carrier as a function of the
interaction duration in the case where the initial state of the
atom is a coherent superposition of neighbouring Wannier-Stark
states. Solid line: $a_n^g(t=0)=a_{n+1}^g(t=0)$ for all $n$.
Dashed line: $a_n^g(t=0)=a_{n+1}^g(t=0)e^{i\pi/2}$ for all $n$.
Dotted line: $a_{-1}^g(t=0)=a_{0}^g(t=0)$ and $a_n^g(t=0)=0$ for
$n\neq -1,0$. The shift is defined as the equilibrium point of a
frequency servo loop using a square frequency modulation of
optimal depth and computed for $U_0=5\,E_r$ and a carrier Rabi
frequency $\Omega_0/2\pi=10\,$Hz. The interaction duration
corresponding to a $\pi$ pulse is
$t\Delta_g/2\pi=43.3$.}\label{fig:coherencestudy}
\end{figure}

The system is more complex if the initial state of the atom is a
coherent superposition of neighbouring wells. In this case
off-resonant excitation of the sidebands will interfere with the
carrier excitation with a relative phase which depends on the
initial relative phase of neighbouring wells and on all the
parameters of the atom-laser interaction ($\Omega$, $\omega$ and
the duration of the interaction). This interference leads to a
modification of the carrier transition probability which is of the
order of $\Omega_1/\Delta_g$ (for the first, and most significant,
sideband). For an interaction close to a $\pi$ pulse, an order of
magnitude of the corresponding carrier pulling is then
$\Omega_1\Omega_0/\Delta_g$ which can be significant. As an
example for $U_0=10\,E_r$ and $\Omega_0/2\pi=10\,$Hz the shift is
about $2\times 10^{-2}\,$Hz, {\it i.e.} several times $10^{-17}$
of the clock transition frequency. This shift is a priori all the
more problematic as the initial atomic state is difficult to know
and control accurately.

\begin{figure}
  \includegraphics[width=5 cm]{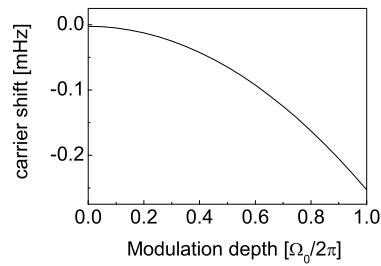}
\caption{Frequency shift of the carrier as a function of the
square modulation depth (see caption of Fig.
\ref{fig:coherencestudy}). The calculation has been performed for
$U_0=5\,E_r$, $\Omega_0/2\pi=10\,$Hz and an interaction time of
$t\Delta_g/2\pi=43.5$. The initial atomic state is the one
corresponding to the dotted line in figure
\ref{fig:coherencestudy}.}\label{fig:coherence_shift_delta}
\end{figure}

We have numerically solved eq. (\ref{eq:aevolutionwannier}) for
various initial atomic states, lattice depths and interaction
parameters to get a more quantitative insight of the effect. The
results are illustrated in figure \ref{fig:coherencestudy} for the
case $U_0=5\,E_r$. A clear signature of the effect can be
identified from its dependence on the interaction duration: the
frequency shift oscillates with a frequency $\Delta_g/2\pi$
resulting from the $\Delta_g$ term in $\Delta_{m-m'}$ in
(\ref{eq:aevolutionwannier}). This provides a powerful method for
investigating site-to-site coherences in the lattice. More
interestingly from a clock point of view, the shift becomes
negligible for all interaction durations $t$ such that
$t=(n+1/2)2\pi/\Delta_g$. For these interaction durations the
interference from the sidebands is symmetric for positive and
negative detunings from resonance, leading to no overall shift.
Since $\Delta_g$ is extremely well known (potentially at the
$10^{-9}$ level) this condition can be accurately met. Note that
choosing such a value of the interaction duration does not
significantly affect the contrast, as the two relevant timescales
have different orders of magnitude in the narrow resonance limit
($\Omega^{-1}>> \Delta_g^{-1}$), and therefore a range of values
of $t$ such that $t=(n+1/2)2\pi/\Delta_g$ correspond to almost
optimal contrast (e.g. all such values of $t$ in figure
\ref{fig:coherencestudy}). A more detailed study shows that the
level of cancellation depends on the depth of the modulation used
to determine the frequency shift (see caption of Fig.
\ref{fig:coherencestudy}) which results from a slight distortion
of the carrier resonance. This effect is shown in figure
\ref{fig:coherence_shift_delta}, which clearly indicates that the
shift can be controlled to below 1 mHz even for a very shallow
lattice depth down to $U_0=5\,E_r$.

\section{Discussion and conclusion\label{sec:conclusion}}

We studied the trap depth requirement for the operation of an
optical lattice clock with a projected fractional accuracy in the
$10^{-17}-10^{-18}$ range. We have shown that using a purely
periodic potential necessitates a lattice of depth $50-100\,E_r$
limited by tunnelling between adjacent sites of the lattice. A
possible way to vastly reduce this depth is to use gravity to lift
the degeneracy between the potential wells which strongly inhibits
tunnelling. Trap depths down to $5-10\,E_r$ are then sufficient to
cancel the effects of the atom dynamics at the desired accuracy
level. This will become even more important for future work aiming
at even higher accuracies. Although very simple, gravity is not
the only way to suppress tunnelling and other solutions,
essentially consisting in a dynamic control of the lattice, are
certainly possible\,\cite{Grossmann91,Diener01,Haroutyunyan01}.
They may prove useful if one wants to operate a lattice clock in
space for instance.

Throughout the paper we have not taken into account the dynamics
of the atoms in the directions transverse to the probe beam
propagation. Experimental imperfections however (misalignement,
wavefront curvature, aberrations) may lead to a residual
sensitivity to this dynamics. If for example the probe beam is
misaligned with respect to the vertical lattice by 100$\,\mu$rad
the transverse wavevector $k_\bot$ is about $10^{-4}\,k_s$ and a
modest transverse confinement should be sufficient to make its
effect negligible. Such a confinement can be provided by the
gaussian transverse shape of the laser forming the lattice or by a
3D lattice. The latter also leads to an interesting physical
problem depending on the relative orientation of the lattice with
respect to gravity\,\cite{Gluck01}.

Finally the well-defined energy separation between Wannier-Stark
states and the possibility to drive transitions between them on
the red or blue sideband of the spectrum (section
\,\ref{sec:results2}) opens new possibilities for the realization
of atom interferometers. This provides a method to generate a
coherent superposition of distant states for the accurate
measurement of the energy separation between these states. This
can for instance lead to an alternative determination of $g$ or
$h/m_a$\,\cite{Weiss94,Gupta02,Battesti04,Modugno04}.

\section*{ACKNOWLEDGEMENTS}
We thank S\'{e}bastien Bize, Andr\'{e} Clairon and Arnaud
Landragin for fruitful and stimulating discussions, as well as
Foss{\'e} Laurent for motivating this work. SYRTE is Unit\'e
Associ\'ee au CNRS (UMR 8630) and acknowledges support from
Laboratoire National de M{\'e}trologie et d'Essai (LNE).


\pagebreak
\end{document}